\documentclass[journal]{IEEEtran}
\usepackage{graphicx}
\usepackage{color}
\usepackage{placeins}
\usepackage{float}
\usepackage{tabularx,colortbl}
\usepackage{amssymb}
\usepackage{amsthm}
\usepackage{cite}
\usepackage{amsmath}

\makeatletter                               % algorithm
\newif\if@restonecol
\makeatother

\makeatletter
\newif\if@restonecol
\makeatother

\usepackage[linesnumbered,ruled,vlined]{algorithm2e}%[ruled,vlined]{
\usepackage{algpseudocode}
\usepackage{amsmath}
\usepackage{listings}                       % Matlab code
\usepackage[framed,numbered,autolinebreaks,useliterate]{mcode}
\theoremstyle{plain}
\newtheorem{thm}{Theorem}[section]

\theoremstyle{plain}
\newtheorem{coro}{Corollary}

\usepackage{amssymb}
\usepackage{bbm}
\usepackage{amsmath,bm}

\IEEEoverridecommandlockouts

\begin{document}
%%%%%%%%%%%%%%%%%%%%%%%%%%%%%%%%%%%%%%%%%%%%%%%%%%%%%%%%%%%%%%%%%%%%%%%
%----------------------------title&author&thanks----------------------------
%\title{Content Caching for Beyond 5G: Cell-Free versus Small Cells}
%%%%%%%%%%%%%%%%%%%%%%%%%%%%%%%%%%%%%%%%%%%%%%%%%%%%%%%%%%%%%%%%%%%%%%%
\title{Uplink Performance of Cell-Free Extremely Large-Scale MIMO Systems}
\author{Hao Lei, Zhe Wang, Huahua Xiao, Jiayi Zhang,~\IEEEmembership{Senior Member,~IEEE}, Bo Ai,~\IEEEmembership{Fellow,~IEEE}\\
\thanks{H. Lei, Z. Wang and J. Zhang are with the School of Electronic and Information
Engineering, Beijing Jiaotong University, Beijing 100044, China, and also
with the Frontiers Science Center for Smart High-speed Railway System,
Beijing Jiaotong University, Beijing 100044, China.

H. Xiao is with ZTE Corporation, State Key Laboratory of Mobile Network and Mobile Multimedia Technology.

B. Ai is with the State Key Laboratory of Rail Traffic Control and Safety,
Beijing Jiaotong University, Beijing 100044, China.
}
}
\maketitle

%\author{Bjtu~Mimo,~\IEEEmembership{Graduate Student Member,~IEEE}\\
%\thanks{B. Mimo is with the School of Electronic and Information Engineering, Beijing Jiaotong University, Beijing 100044, China.
%(e-mail: 18110000@bjtu.edu.cn).}
%}

%%%%%%%%%%%%%%%%%%%%%%%%%%%%%%%%%%%%%%%%%%%%%%%%%%%%%%%%%%%%%%%%%%%%%%%
%----------------------------abstract----------------------------
%%%%%%%%%%%%%%%%%%%%%%%%%%%%%%%%%%%%%%%%%%%%%%%%%%%%%%%%%%%%%%%%%%%%%%%
\begin{abstract}
In this paper, we investigate the uplink performance of cell-free (CF) extremely large-scale multiple-input-multiple-output (XL-MIMO) systems, which is a promising technique for future wireless communications.
More specifically, we consider the practical scenario with multiple base stations (BSs) and multiple user equipments (UEs).
%Then, two different uplink processing schemes for the XL-MIMO called the CF XL-MIMO and the small-cell XL-MIMO are implemented.
To this end, we derive exact achievable spectral efficiency (SE) expressions for any combining scheme.
It is worth noting that we derive the closed-form SE expressions for the CF XL-MIMO with maximum ratio (MR) combining.
Numerical results show that the SE performance of the CF XL-MIMO can be hugely improved compared with the small-cell XL-MIMO.
It is interesting that a smaller antenna spacing  leads to a higher correlation level among patch antennas.
Finally, we prove that increasing the number of  UE antennas may decrease the SE performance with MR combining.
\end{abstract}

%%%%%%%%%%%%%%%%%%%%%%%%%%%%%%%%%%%%%%%%%%%%%%%%%%%%%%%%%%%%%%%%%%%%%%%
%----------------------------keywords----------------------------
%%%%%%%%%%%%%%%%%%%%%%%%%%%%%%%%%%%%%%%%%%%%%%%%%%%%%%%%%%%%%%%%%%%%%%%

%\begin{IEEEkeywords}
%
%Extremely large-scale multiple-input-multiple-output, spectral efficiency, near-filed communications.
%
%\end{IEEEkeywords}

%\newpage
\IEEEpeerreviewmaketitle
\vspace{-0.3cm}
%%%%%%%%%%%%%%%%%%%%%%%%%%%%%%%%%%%%%%%%%%%%%%%%%%%%%%%%%%%%%%%%%%%%%%%
%----------------------------introduction----------------------------
%%%%%%%%%%%%%%%%%%%%%%%%%%%%%%%%%%%%%%%%%%%%%%%%%%%%%%%%%%%%%%%%%%%%%%%
\section{Introduction}

The extremely large-scale multiple-input-multiple-output (XL-MIMO) is a promising technique for the future communication system where the system is equipped with a massive number of antennas in a compact space \cite{bjornson2019massive}, \cite{9113273}. Compared with the conventional massive MIMO (mMIMO), the XL-MIMO can provide higher spectral efficiency (SE) and higher energy efficiency (EE) with densely deployed antennas. Specifically, there are many hardware design schemes for realizing the XL-MIMO system, such as holographic MIMO, large intelligent surfaces (LIS), extremely large antenna array (ELAA), and continuous aperture MIMO (CAP-MIMO) \cite{https://doi.org/10.48550/arxiv.2209.12131}.
However, the electromagnetic (EM) characteristics of the XL-MIMO are quite different. %\cite{https://doi.org/10.48550/arxiv.2209.12131}, \cite{9903389}.
The commonly used uniform plane wave (UPW) model with far-field assumptions may fail in the XL-MIMO.
We should consider the scenario that the UEs are located in the near-field region and the EM channel has to be modeled by spherical wavefront \cite{https://doi.org/10.48550/arxiv.2209.12131}, \cite{9903389}.
Moreover, the effective antenna areas and the losses from polarization mismatch should also be considered \cite{https://doi.org/10.48550/arxiv.2209.03082}.

Recently, the effect of the near-field characteristics has been initially considered in the XL-MIMO  systems \cite{https://doi.org/10.48550/arxiv.2209.12131}, \cite{9903389}, [6]-[12].
In \cite{https://doi.org/10.48550/arxiv.2209.12131}, the authors comprehensively reviewed the existing XL-MIMO hardware designs, discussed the existing challenges and proposed some promising solutions for the XL-MIMO.
In \cite{9903389}, the non-linear phase property of spherical waves was introduced and the metric was proposed to determine the near-field ranges in typical communication scenarios.
The  authors of \cite{9650519} proposed a EM channel matrix deduced from the dyadic Green's function and discussed the the degree of freedom (DoF) and effective degree of freedom (EDoF) limit of the XL-MIMO.
In \cite{9136592}, densely packed sub-wavelength patch antennas are incorporated to approximately realize a spatially-continuous aperture.
The authors of  \cite{9443506}, \cite{9110848} considered the small-scale fading and provided a plane-wave representation of the XL-MIMO from an electromagnetic perspective.
A channel model different from \cite{9110848} by considering non-isotropic scattering and directive antennas was  developed by the authors of \cite{9716880}.
In \cite{9724113}, the authors proposed a novel Fourier plane-wave stochastic scalar channel model for single-user XL-MIMO communication systems by fully capturing the essence of electromagnetic propagation. Moreover, the authors of \cite{9779586} extended the scenario in
\cite{9724113} to multi-user communication systems and analyzed the SE performance of the XL-MIMO.

As a promising technique for the future wireless communication, cell-free (CF) mMIMO  has been proven to achieve higher SE by deploying a large number of distributed access points, compared with cellular mMIMO \cite{[1]}.
%\cite{[3]} \cite{[4]} \cite{[5]} \cite{[6]} \cite{[7]} \cite{[8]} \cite{[9]} \cite{[10]}
However, practical scenarios with multiple BSs are rarely considered in the XL-MIMO.
More importantly, although the system with a large number of distributed BSs benefits from the macro diversity, its interference suppression ability depends on how it operates.
And the cooperation among multiple XL-MIMO BSs is not discussed.
Besides, the signal processing for the XL-MIMO displays high complexity due to the extremely large-scale antennas.
Therefore, the signal processing scheme with acceptable complexity is also a challenge.

Motivated by the EM channel model of \cite{9724113}, \cite{9779586}, this paper introduce a novel XL-MIMO channel model for the scenario with  multi-BS multi-UE. Inspired by the promising cell-free mMIMO technology \cite{9113273}, \cite{9174860}, we investigate two different uplink processing schemes
of the XL-MIMO, called the CF XL-MIMO and the small-cell XL-MIMO.
The major contributions of this paper are as follows:
\begin{itemize}
  \item We extend the channel model of single-BS multi-UE in \cite{9779586} to the one of multi-BS multi-UE in the XL-MIMO. More importantly, we investigate the uplink performance of both the CF XL-MIMO and the small-cell XL-MIMO to reveal the benefits of the distributed network.
  \item We derive exact achievable uplink SE expressions for two signal processing schemes with arbitrary combining scheme. Moreover, we propose closed-form SE expressions for the CF XL-MIMO with MR combining\footnote{ Simulation codes are provided to reproduce the results in this paper: https://github.com/BJTU-MIMO.}.
%  \item We reveal the significant impact of antenna spacing on the uplink performance of the CF XL-MIMO and provide physical insights for the practical CF XL-MIMO implementation.
\end{itemize}

%{\textbf{\textit{Notation}}:}
%Boldface lowercase letters $\rm {\bf{a}}$ and boldface uppercase letters $\rm {\bf{A}}$ denote column vectors and matrices, respectively. Conjugate,
%transpose, and conjugate transpose are denoted by $(\cdot)^{*}$, $(\cdot)^{T}$ and $(\cdot)^{H}$. $\rm{diag}(\rm {\bf{a}})$ a diagonal matrix with the entries of $\rm {\bf{a}}$ on the diagonal. We use $\mathbb{E} \{ \cdot \} $ and $\rm{tr}\{ \cdot \} $ to denote the expectation operator and the trace operator, respectively. The definitions and the determinant of a matrix are denoted
%by $ \triangleq$ and $ \left| \cdot \right| $, respectively. $\otimes$ and $\odot$ denote the Kronecker products and the element-wise products, respectively. ${\rm \bf{I}}_n $ is the $n\times n$ identity matrix,
%and $ {\rm \bf{1}}_n$ is a column vector with all ones. The circularly symmetric complex Gaussian distribution is denoted by $\mathcal{CN}(0,\sigma^2)$.

\begin{figure}
\centering
\includegraphics[width=2.7in]{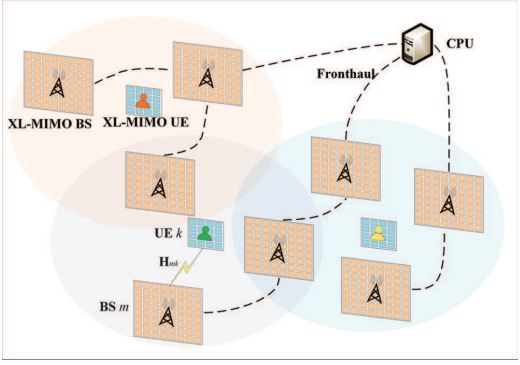}
\caption{Illustration of the CF XL-MIMO system}
\end{figure}

%%%%%%%%%%%%%%%%%%%%%%%%%%%%%%%%%%%%%%%%%%%%%%%%%%%%%%%%%%%%%%%%%%%%%%%
%-----------------------system and channel modeling--------------------
%%%%%%%%%%%%%%%%%%%%%%%%%%%%%%%%%%%%%%%%%%%%%%%%%%%%%%%%%%%%%%%%%%%%%%
\section{System Model}

As illustrated in Fig. 1, the uplink of a CF XL-MIMO network is investigated where $M$ BSs and $K$ UEs are arbitrarily distributed in a wide area. To reduce the computation overhead, the BSs are connected to a central processing unit (CPU) with high computation processing
ability by fronthaul links.
Each BS consists of a planar extremely large-scale surface (XL-surface) with $N_{r} = N_{H_r}N_{V_r}$ patch antennas where $N_{H_r}$ and $N_{V_r}$ denote the number of antennas per row and per column, respectively.
The horizontal and vertical patch antenna spacing $\Delta_{r}  $ is below $ {\lambda  \mathord{\left/
 {\vphantom {\lambda  2}} \right. \kern-\nulldelimiterspace} 2}$.
 We denote the horizontal and vertical length of the XL-surface by $L_{r,x}=N_{Hr}\Delta_{r} $ and $L_{r,y}=N_{Vr}\Delta_{r} $, respectively. The antennas at each BS are indexed row-by-row by $ n \in [1,N_{r}]$. The location of the $m$-th BS with respect to the origin is $ {\rm{\bf r}}_{m} = [ r_{m,x},r_{m,y},r_{m,z}  ]^{T}  $. The receive response vector is denoted as $ {{{ \rm{\bf a}}}_r}( {\bf k} ,{\rm{{\bf r}}}) \!\!\!=\!\!\! \left[ {{{\bf a}_{r,1}}({\bf k} ,{\rm{{\bf r}}}), \ldots ,{{\bf a}_{r,{M}}}({\bf k} ,{\rm{\bf r}})} \right] \!\! \in\!\! \mathbb{C}{^{M \times 1}} $ with
 \begin{equation}\label{eq:one}
 {{\bf a}_{r,{m}}}({\bf k} ,{\rm{\bf r}}) = \left[ e^{j{\bf k} {{\left( {  \varphi ,\theta } \right)}^T} {{ \rm{ {\bf r}} }_1^{(m)} }}, \ldots ,e^{j{\bf k} {{\left( {\varphi ,\theta } \right)}^T}{{ \rm{{\bf r}}}_{N_r}^{(m)}}} \right]^T ,
\end{equation}
where ${ \bf{k}} \left( {  \varphi_r , \theta_r } \right)=[k_x,k_y,k_z] = [k {\rm cos}(\theta_r) {\rm cos}(\varphi_r), k {\rm cos}(\theta_r) {\rm sin}(\varphi_r), {\rm sin}(\theta_r)] $ is the receive wave vector with the wavenumber $ k = {{2\pi } \mathord{\left/
 {\vphantom {{2\pi } \lambda }} \right. \kern-\nulldelimiterspace} \lambda } $, the receive azimuth angle $ \varphi_r $ and the receive elevation angle $  \theta_r $. And $ {{ \rm{{\bf r}}}_{n}^{(m)}} $ is the location of the $n$-th antenna of $m$-th BS.

Similarly, each UE is equipped with a planar XL-surface with $N_{s}$ antennas.
% Similarly, each UE is equipped with a planar XL-surface with $N_{s}$ patch antennas.
And we assume that all planar XL-surface are parallel.
The horizontal and vertical antenna spacing, the horizontal length and the vertical length are denoted by $\Delta_{s}   $, $ L_{s,x} $ and $L_{s,y} $, respectively.
We denote the location of $k$-th user by $ {\rm{\bf s}}_{k} = [ s_{k,x},s_{k,y},s_{k,z}  ]^{T} , k = 1,...,K$. The transmit wave vector is denoted as ${{\rm{{\bf a}}}_s}( {\bm {\kappa}}  ,{\rm{{\bf s}}}) \!\!=\!\! \left[ {{{\rm{{\bf a}}}_{s,1}}({\bm \kappa} ,{\rm{{\bf s}}}), \ldots ,{{\rm{{\bf a}}}_{s,{K}}}( {\bm \kappa} ,{\rm{{\bf s}}})}  \right] $ with
\begin{equation}
\label{eq:two}
{{\rm{{\bf a}}}_{s,k}}\left( {{\bm \kappa} ,{\rm{{\bf s}}}} \right) = \left[ e^{j{\bm \kappa} ^T{\rm{{\bf s}}}_1^{(k)}}, \ldots ,e^{j {\bm \kappa}^T{\rm{{\bf s}}}_{N_s}^{(k)}} \right]^T,k=1,...,K,
\end{equation}
where ${\bm \kappa}  \in \mathbb{C}{^{3}}  $ is the transmit wave vector ${\bm \kappa} = [\kappa_x,\kappa_y,\kappa_z] = [k {\rm cos}(\theta_s) {\rm cos}(\varphi_s), k {\rm cos}(\theta_s) {\rm sin}(\varphi_s), {\rm sin}(\theta_s)] $ with the  transmit azimuth angle $ \varphi_s $ and the transmit elevation angle $  \theta_s $.

%%%%%%%%%%%%%%%%%%%%%%%%%%%%%%%%%%%%%%%%%%%%%%%%%%%%%%%%%%%%%%%%%%%%%%%
%-----------------------Existing Channel Modeling--------------------
%%%%%%%%%%%%%%%%%%%%%%%%%%%%%%%%%%%%%%%%%%%%%%%%%%%%%%%%%%%%%%%%%%%%%%
\subsection{Channel Modeling for the Single-BS Single-UE Scenario}

%%%%%%%%%%%%%%%%%%%%%%%%%%%%%%%%%%%%%%%%%%%%%%%%%%%%%%%%%%%%%%%%%%%%%%%
%-----------------------Individual Users--------------------
%%%%%%%%%%%%%%%%%%%%%%%%%%%%%%%%%%%%%%%%%%%%%%%%%%%%%%%%%%%%%%%%%%%%%%%
%\subsubsection{Individual Users}

In the XL-MIMO communication system, the near-field region is extended to tens of meters, which resluts in UEs more likely to be in the near-field.
Then the EM channel should be accurately modeled based on the spherical wave assumption.
%In \cite{9110848}, \cite{9724113}, the stochastic model was utilized to depict the electromagnetic non-line-of-sight (NLoS) propagation  under arbitrary scattering conditions in the near-filed.
%Starting from Maxwell's equations, the authors of \cite{9724113} proposed a Fourier plane-wave series expansion of the channel response based on the scalar Green's function.

Specifically, by considering the scenario with single-BS single-UE, the $ (m,n)$-entry of the space domain channel $ {{\rm{\bf H}}}  \in \mathbb{C}^{N_r\times N_s} $ can be denoted as \cite{9724113}
\begin{equation}
\begin{aligned}\label{eq:three}
{[{\rm{\bf H}}]_{mn}}\! =\! \frac{1}{{{{\left( {2\pi } \right)}^2}}}\iiiint_{ \mathcal{D}\times  \mathcal{D}}&\! {{a_{r,m}}\left( {{\rm{\bf k}},{\rm{\bf r}}}\right)\!{H_a}\!\left( {{k_x},{k_y},{\kappa _x},{\kappa _y}} \right)} \\
&{a_{s,n}}\left( {{\bm \kappa} ,{\rm{\bf s}}} \right)d{k_x}d{k_y}d{\kappa _x}d{\kappa _y},
\end{aligned}
\end{equation}
where ${a_{r,m}}\left( {{\rm{\bf k}},{\rm{\bf r}}}\right)$ is the $m$-th element in \eqref{eq:one}, ${a_{s,n}}\left( {{\bm \kappa} ,{\rm{\bf s}}}\right)$ is the $n$-th element in \eqref{eq:two}, $ {H_a} \left( {{k_x},{k_y},{\kappa _x},{\kappa _y}} \right)  $ is the wavenumber domain channel, and the integration region is  $ \mathcal{D}=\left\{ (k_x,k_y) \in \mathbb{C}{^{2}} : k_x^2+k_y^2 \leq \kappa^2 \right\} $, respectively.

\begin{figure*}[!t]
\normalsize
\setcounter{equation}{11}
\begin{equation}\label{eq:thirteen}
{{\rm{ \bf R}}_{mk}} = \left( {{{\left( {{\rm{ \bf U}}_s^{\left( k \right)}} \right)}^*} \otimes {\rm{ \bf U}}_r^{\left( m \right)}} \right)
\left( {{\rm diag}\left( {{\bm \sigma} _k^{(s)} \odot  {\bm \sigma} _k^{(s)}} \right) \otimes {\rm diag}\left( {{\bm \sigma} _m^{(r)} \odot {\bm \sigma} _m^{(r)}} \right)} \right)
\left( {{{\left( {{\rm{ \bf U}}_s^{\left( k \right)}} \right)}^T} \otimes {{\left( {{\rm{ \bf U}}_r^{\left( m \right)}} \right)}^H}} \right)
\end{equation}
\hrulefill
\vspace*{0.5pt}
\end{figure*}

%The spatial domain channel can be given by the wavenumber domain channel. And the transformation between the space domain and the wavenumber domain is similar to the Fourier transformation between the time domain and the frequency domain.

As shown in \cite{9724113}, the wavenumber domain channel in \eqref{eq:three} can be denoted as
\begin{equation}
\begin{aligned}
\setcounter{equation}{4}
\!\!&{H_a} \!\left(\! {{k_x},{k_y},{\kappa _x},{\kappa _y}} \!\right)\! =\!  {S^{1/2}}\!\left(\! {{k_x},{k_y},{\kappa _x},{\kappa _y}} \!\right)\!W\!\left(\! {{k_x},{k_y},{\kappa _x},{\kappa _y}} \!\right) \!\!\!
\\
 &=\! \frac{{\kappa \eta }}{2}\!\frac{{A\!\left( {{k_x},{k_y},{\kappa _x},{\kappa _y}} \right)\!W\!\left( {{k_x},{k_y},{\kappa _x},{\kappa _y}} \right)}}{{k_z^{1/2}\left( {{k_x},{k_y}} \right)\kappa _z^{1/2}\left( {{\kappa _x},{\kappa _y}} \right)}},
\end{aligned}
\end{equation}
where $ S\left( {{k_x},{k_y},{\kappa _x},{\kappa _y}} \right) = \frac{{{A^2}\left( {{k_x},{k_y},{\kappa _x},{\kappa _y}} \right)}}{{k_z\left( {{k_x},{k_y}} \right)\kappa _z\left( {{\kappa _x},{\kappa _y}} \right)}}$ is the spectral density, $ {A}\left( {{k_x},{k_y},{\kappa _x},{\kappa _y}} \right) $ is an arbitrary real-valued, non-negative function and $W\! \left( {{k_x},{k_y},{\kappa _x},{\kappa _y}} \right) \!\!\!\!\sim\! \!\!\mathcal{CN}(0,1) $ is a collection of random characteristics. In the isotropic
scattering condition, we have $ A\left( {{k_x},{k_y},{k _z}} \right) \!\!\!=\!\!\!\frac{2\pi}{\sqrt{k}} $ with unit average power.

The wavenumber domain channel displays sparse characteristics where only finite elements are non-zero. Thus, the space domain channel in \eqref{eq:three} can be  approximated with finite sampling points within the lattice ellipse \cite{9724113}, \cite{9779586}
\begin{equation}
\begin{aligned}
{\mathcal{E}_s} \!&= \!\left\{\! {\left( {{m_x},{m_y}} \right)\! \in\! {\mathbb{Z}^2}\!:\!{{\left( {{m_x}\lambda /{L_{s,x}}} \right)}^2}\! +\! {{\left( {{m_y}\lambda /{L_{s,y}}} \right)}^2}\! \le \!1} \!\right\},\!\!\!
\\
{\mathcal{E} _r}\! &= \!\left\{\! {\left( {{\ell _x},{\ell _y}} \right) \in {\mathbb{Z}^2}:{{\left( {{\ell _x}\lambda /{L_{r,x}}} \right)}^2} + {{\left( {{\ell _y}\lambda /{L_{r,y}}} \right)}^2} \le 1} \right\},
\end{aligned}
\end{equation}
at each UE and each BS, respectively. And we denote the cardinalities of the sets $ {{\mathcal{E}_s}}   $ and  $ {{\mathcal{E}_r}}  $  by $ n_s= \left| {{\mathcal{E}_s}} \right|  $ and $ n_r= \left| {{\mathcal{E}_r}} \right|  $, respectively.

The channel ${\rm{\bf H}}$ in \eqref{eq:three} can be approximately described by a 4D Fourier plane-wave series expansion \cite{9724113}
\begin{equation}
\begin{aligned}
{\left[{\rm{\bf H}} \right]_{mn}} \approx \sum\limits_{\left( {{l_x},{l_y}} \right) \in {\varepsilon _r}} \sum\limits_{\left( {{m_x},{m_y}} \right) \in {\varepsilon _s}} {H_a}\left( {{\ell _x},{\ell _y},{m_x},{m_y}} \right)
\\
{a_{r,m}}\left( {{\ell _x},{\ell _y},{\rm{{\bf r}}}} \right){a_{s,n}}\left( {{m_x},{m_y},{\rm{{\bf s}}}} \right),
\end{aligned}
\end{equation}
with Fourier coefficients
\begin{equation}
{H_a}\left( {{\ell _x},{\ell _y},{m_x},{m_y}} \right) \sim \mathcal{CN}(0, \sigma^2\left( {{\ell _x},{\ell _y},{m_x},{m_y}} \right)  ).
\end{equation}
The variance $  \sigma^2\left( {{\ell _x},{\ell _y},{m_x},{m_y}} \right) $ is the variance of sampling point $ \left( {{\ell _x},{\ell _y},{m_x},{m_y}} \right) $, which can be formulated under the separable scattering shown as  \cite{9724113}, \cite{9779586}
\begin{equation}
\begin{aligned}
{\sigma ^2} & \left( {{\ell _x},{\ell _y},{m_x},{m_y}} \right) \propto \iiiint_{
{{{\hat S}_s} \times {{\hat S}_r}}}\!\!\!\!\!\!\!  {
{{\mathbbm{1}_{\hat {\mathcal{D}}}}\left( {{k_x},{k_y}} \right){\mathbbm{1}_{\hat {\mathcal{D}}}}\left( {{\kappa _x},{\kappa _y}} \right)}} \!\!
\\
&\frac{{{A^2}\left( {{{\hat k}_x},{{\hat k}_y},{{\hat \kappa }_x},{{\hat \kappa }_y}} \right)}}{{{{\hat k}_z}\left( {{{\hat k}_x},{{\hat k}_y}} \right){{\hat \kappa }_z}\left( {{{\hat \kappa }_x},{{\hat \kappa }_y}} \right)}}d{{\hat k}_x}d{{\hat k}_y}d{{\hat \kappa }_x}d{{\hat \kappa }_y} \!\!\!\!\!\!\!\!
\\
 &= \sum\limits_{i = 1}^{i = 3} {\sum\limits_{j = 1}^{j = 3} {\iiiint_{
{{\Omega _{s,j}}\left( {{m_x},{m_y}} \right) \times {\Omega _{r,i}}\left( {{\ell _x},{\ell _y}} \right)} } \!\!\!\!\!\!\!\!\!\!\!\!\!\!\!\!\!\!\!\!\!\!\!\!\!\!\!\!\!\!\!\!\!\! \!\!\!\!\! {{A^2}\left( {{\theta _r},{\phi _r},{\theta _s},{\phi _s}} \right)d{\Omega _s}d{\Omega _r}}     } } ,
\end{aligned}
\end{equation}
where $  {{\hat k}_x},{{\hat k}_y},{{\hat k}_z} = (k_x,k_y,k_z)/k    $ are normalized wavevector coordinates, and $ {\Omega _{r}}\left( {{\ell _x},{\ell _y}} \right)$ and $ {\Omega _{s}}\left( {{m_x},{m_y}}\right) $ are integration regions in the wavenumber domain of each BS and each UE, respectively \cite[ Appendix IV.C]{9110848}.

\begin{figure*}[!t]
\normalsize
\setcounter{equation}{17}
\begin{equation}\label{eq:twenty}
{\rm SE}_{mk}^{(2)} =  {\mathbb E}\left\{ {\log _2} \left|  {\rm {\bf I}}_{N_S}   +   p {   {{\rm {\bf H}}_{mk}^H{{\rm {\bf H}}_{mk}}} }
\left(   p\sum_{l = 1,l \neq k}^K { {  {{\rm {\bf H}}_{mk}^H{{\rm {\bf H}}_{ml}}{\rm {\bf H}}_{ml}^H{{\rm {\bf H}}_{mk}}} } } + \sigma_w^2 {  {{\rm {\bf H}}_{mk}^H{{\rm {\bf H}}_{mk}}} }
 \right)^{-1}
 {   {{\rm {\bf H}}_{mk}^H{{\rm {\bf H}}_{mk}}} }
 \right|\right\}
\end{equation}
\hrulefill
\vspace*{0.5pt}
\end{figure*}

%%%%%%%%%%%%%%%%%%%%%%%%%%%%%%%%%%%%%%%%%%%%%%%%%%%%%%%%%%%%%%%%%%%%%%%
%-----------------------Channel Modeling for XL-MIMO--------------------
%%%%%%%%%%%%%%%%%%%%%%%%%%%%%%%%%%%%%%%%%%%%%%%%%%%%%%%%%%%%%%%%%%%%%%
\subsection{Channel Modeling for the Multi-BS Multi-UE Scenario}
In \cite{9779586}, the authors extended  the scenario with single-BS single-UE channel modeling from
the previous subsection to the scenario with single-BS multi-UE. Similarly,
the channel between $m$-th BS and $k$-th UE $ {\rm{ \bf H}}_{mk}\in \mathbb{C}^{N_r\times N_s} $ is
\begin{equation}
\begin{aligned}
\setcounter{equation}{9}
\label{eq:ninelh}
{\rm{ \bf H}}_{mk}\! =\! &\sqrt {{N_r}{N_s}}\!\!\!\! \sum\limits_{\left( {{\ell _x},{\ell _y}} \right) \in {\mathcal{E} _r}} {\sum\limits_{\left( {{m_x},{m_y}} \right) \in {\mathcal{E} _s}}\!\!\! {H_a^{\left( {mk} \right)}\left( {{\ell _x},{\ell _y},{m_x},{m_y}} \right)} }\!\!\!\!\!\!
 \\
&{{\rm{{\bf a}}}_r}\left( {{\ell _x},{\ell _y},{{\rm{{\bf r}}}^{\left( m \right)}}} \right){{\rm{{\bf a}}}_s}\left( {{m_x},{m_y},{{\rm{{\bf s}}}^{\left( k\right)}}} \right),
\end{aligned}
\end{equation}
where the Fourier coefficient
\begin{equation}
\!\!{H_a^{\left( {mk} \right)}\left( {{\ell _x},{\ell _y},{m_x},{m_y}} \right)} \sim \mathcal{N}_{\mathbb{C}}( 0, \sigma_{mk}^2( {{\ell _x},{\ell _y},{m_x},{m_y}} )),
\end{equation}
and
\begin{equation}
\begin{aligned}
\!\!\!\!\!{\left[ {{{\rm{a}}_s}\!\!\left( \! {{m_x},\! {m_y},\! {{\rm{{\bf s}}}^{\left( k \right)}}}\! \right)}\! \right]_j} \!\! \!\!  &=\!\! \frac{1}{{\sqrt {{N_s}} }}{e^{ \!-\!\!  j\!\left(\! {\frac{{2\pi }}{{{L_{s,\!x}}}} \! {m_x}s{{_x^{\left( k \right)}}\!\!\!\! _{_j}} +\! \frac{{2\pi }}{{{L_{s,\!y}}}}{m_y}s{{_y^{\left( k \right)}}\!\!\!\!_{_j}} + {\gamma _s}\!\left(\! {{m_x},{m_y}}\! \right)s{{_z^{\left( k \right)}}\!\!\!\!_{_j}}} \right)}},\\
j &= 1, \ldots ,{N_s},\\
\!\!{\left[ {{{\rm{a}}_r}\!\!\left( {{\ell _x},{\ell _y},{{\rm{{\bf r}}}^{\left( m \right)}}} \!\right)}\! \right]_i}\!\! \!   &=\!\! \frac{1}{{\sqrt {{N_r}} }}{e^{\! j\! \left( {\frac{{2\pi }}{{{L_{r,x}}}}{\ell _x}r{{_x^{\left( m \right)}}\!\!\!\!_{_i}} + \frac{{2\pi }}{{{L_{r,y}}}}{\ell _y}r{{_y^{\left( m \right)}}\!\!\!\!_{_i}} + {\gamma _r}\left( {{\ell _x},{\ell _y}} \right)r{{_z^{\left( m \right)}}\!\!\!\!_{_i}}} \right)}},\\
i &= 1, \ldots ,{N_r}.
\end{aligned}
\end{equation}

Inspired by \cite{9779586}, the channel in (9) can be written as
$
 {\rm{ \bf H}}_{mk} = {\rm{ \bf U}}_r^{\left( m \right)}{\rm{ \bf H}}_a^{\left( {mk} \right)}{\left( {{\rm{ \bf U}}_s^{\left( k \right)}} \right)^H}
  = {\rm{ \bf U}}_r^{\left( m \right)} \left( {{{\bm\Sigma} ^{\left( {mk} \right)}}\odot {{\rm{ \bf W}}^{\left( {mk} \right)}}}\right)  {\left( {{\rm{ \bf U}}_s^{\left( k \right)}} \right)^H}
$,
where $  {\rm{ \bf U}}_r^{\left( m \right)} $ and $ {{\rm{ \bf U}}_s^{\left( k \right)}}  $ denote the deterministic matrices collecting the variances of $ n_r $ and $ n_s$ sampling points in $ \left( {{\ell _x},{\ell _y},{{\rm{{\bf r}}}^{\left( m \right)}}} \right) $ and $ \left(  {{m_x}, {m_y}, {{\rm{{\bf s}}}^{\left( k \right)}}} \right) $, respectively. And $ {\rm{ \bf H}}_a^{\left( {mk} \right)} =   {{{\bm\Sigma} ^{\left( {mk} \right)}}\odot {{\rm{ \bf W}}^{\left( {mk} \right)}}}  \in \mathbb{C}^{n_r\times n_s} $ collects $ {\sqrt {{N_r}{N_s}} H_a^{\left( {mk} \right)}\left( {{\ell _x},{\ell _y},{m_x},{m_y}} \right)} $ for $ n_r \!\cdot n_s $ sampling points, and $ {\rm{ \bf W}} \sim \mathcal{CN}( 0, {\rm{{\bf I}}}_{n_r } ) $. We have $ {{\bm\Sigma} ^{\left( {mk} \right)}} = ({\bm \sigma} _m^{(r)}  {\rm \bf{1}}_{n_s}^T )\odot ({\rm \bf{1}}_{n_r} ({\bm \sigma} _k^{(s)} )^T ) $ where $ {\bm \sigma} _m^{(r)} \in \mathbb{R}^{n_r \times 1} $ and $ {\bm \sigma} _k^{(s)} \in  \mathbb{R}^{n_s \times 1}$ collect $  \sqrt {{N_r}}  { \sigma} _m^{(r)}\left( {{\ell _x},{\ell _y}} \right) $ and $  \sqrt {{N_s}}  { \sigma} _k^{(s)}\left( {{m _x},{m _y}} \right) $, respectively.

As in \cite{9737367}, we can structure the channel as $ {\rm{\bf h}}_{mk}={\rm vec}(  {\rm{ \bf H}}_{mk} ) \sim \mathcal{N}_{\mathbb{C}}(0,   {\rm{ \bf R}}_{mk})$, where $ {\rm{ \bf R}}_{mk} \triangleq \mathbb{E} \left\{ {\rm vec}( {\rm{ \bf H}}_{mk} ){\rm vec}( {\rm{ \bf H}}_{mk} )^H \right\} $ is the full correlation matrix as \eqref{eq:thirteen}, shown at the top of this page. The full correlation matrix can also be structured in the block form as \cite{9737367} where $  {{\rm{ \bf R}}_{mk}^{ni}} = \mathbb{E}\{  { \rm { \bf h}}_{mk,n} { \rm { \bf h}}_{mk,i}^H \}  $ with ${ \rm { \bf h}}_{mk,n} $ and $ { \rm { \bf h}}_{mk,i}$ being the $n$-th column and $i$-th column of $  {\rm{ \bf H}}_{mk} $, respectively.
%%%%%%%%%%%%%%%%%%%%%%%%%%%%%%%%%%%%%%%%%%%%%%%%%%%%%%%%%%%%%%%%%%%%%%%
%-----------------------Uplink Data Transmission--------------------
%%%%%%%%%%%%%%%%%%%%%%%%%%%%%%%%%%%%%%%%%%%%%%%%%%%%%%%%%%%%%%%%%%%%%%
\subsection{Uplink Data Transmission }

During the uplink, all $K$ UEs sent their data to the BSs. The transmitted signal from UE $k$ is denoted by $ {{\rm {\bf x}}_k}=[ x_{k,1},\cdots ,  x_{k,N_s}]^T  \in \mathbb{C}^{N_s}$ where $  {\rm tr} (  {{\rm {\bf x}}_k}{{\rm {\bf x}}_k^H}  ) = p $ with $p$ being the transmitted power.
The received signal $ {\rm{\bf y}}_m \in \mathbb{C}^{N_r}$ at BS $m$ is
${\rm{\bf y}}_m = \sum_{k = 1}^K { {{\rm{ \bf H}}_{mk}}{{\rm {\bf x}}_k}}  + {{\rm {\bf n}}_m}$,
where $ {{\rm {\bf n}}_m} \sim \mathcal{CN}(0,\sigma_w^2 {\rm{\bf I}}_{N_r})$ is the independent receiver noise with $ \sigma_w^2  $ being the noise power.

%%%%%%%%%%%%%%%%%%%%%%%%%%%%%%%%%%%%%%%%%%%%%%%%%%%%%%%%%%%%%%%%%%%%%%
%-----------------------FOUR SIGNAL PROCESSING SCHEMES--------------------
%%%%%%%%%%%%%%%%%%%%%%%%%%%%%%%%%%%%%%%%%%%%%%%%%%%%%%%%%%%%%%%%%%%%%
\section{Signal Processing Schemes}
%In this section, we consider two different uplink signal processing schemes for the XL-MIMO, called the CF XL-MIMO and the small-cell XL-MIMO. And we derive the achievable SE expressions for two signal processing schemes.

%%%%%%%%%%%%%%%%%%%%%%%%%%%%%%%%%%%%%%%%%%%%%%%%%%%%%%%%%%%%%%%%%%%%%%%
%------Level 2: Local Processing & Simple Centralized Decoding-----------
%%%%%%%%%%%%%%%%%%%%%%%%%%%%%%%%%%%%%%%%%%%%%%%%%%%%%%%%%%%%%%%%%%%%%%
\subsection{Cell-Free XL-MIMO}

In the CF XL-MIMO, the BSs firstly decode the received signals and then transmit the local processed signals to the central processing unit (CPU) for further processing \cite{https://doi.org/10.48550/arxiv.2209.12131}.
%Then the computation overhead of BSs is reduced in the distributed XL-MIMO.

Let $ {\rm {\bf V}}_{mk} \in \mathbb{C}^{N_r \times N_s}$ denote the combining matrix designed by BS $m$ for UE $ k$. Then the local estimate of ${{\rm {\bf x}}_k}$ at BS $m$ is
\begin{equation}
\begin{aligned}
\setcounter{equation}{13}
\label{eq:fourteen}
\!\!\!\!\!\!{{{\rm{\check { {\bf x}}}}}_{mk}} \!\!=\!\! {\rm {\bf V}}_{mk}^H{{\rm {\bf y}}_m}
 \!\!=\!\! {\rm {\bf V}}_{mk}^H{{\rm {\bf H}}_{mk}}{{\rm{{{\bf x}}}}_k} \!\!+\!\!\!\!\!\! \sum\limits_{l = 1,l \ne k}^K \!\!\!\!\! {{\rm {\bf V}}_{mk}^H{{\rm {\bf H}}_{ml}}{{\rm{{ {\bf x}}}}_l}}  \!\!+\!\! {\rm {\bf V}}_{mk}^H{{\rm {\bf n}}_m}.
 \end{aligned}
\end{equation}
Then the local estimates $ {{{\rm{\check { {\bf x}}}}}_{mk}}   $ are sent to the CPU where they are weighted evenly as $ {{{\rm{\hat {\bf x}}}}_k} = \sum_{m = 1}^M { \frac{1}{M}    {{{\rm{\check { {\bf x}}}}}_{mk}}   }
$. The final estimate of ${{\rm {\bf x}}_k}$ at the CPU is denoted by
\begin{equation}
\begin{aligned}
 \!\!{{{\rm{\hat {\bf x}}}}_k}  \!=\! \frac{1}{M} \sum\limits_{m = 1}^M \!{{\rm {\bf V}}_{mk}^H{{\rm {\bf H}}_{mk}}{{\rm{{{\bf x}}}}_k}} \!\! +\!\! \frac{{1 }}{M}\sum\limits_{m = 1}^M {\sum\limits_{l = 1,l \ne k}^K \!\!\!{{\rm {\bf V}}_{mk}^H{{\rm {\bf H}}_{ml}}{{\rm{{{\bf x}}}}_l}} }\!  +\! {{\rm {\bf n'}}\!_k},
 \end{aligned}
\end{equation}
with ${{\rm {\bf n'}}\!_k} =\frac{1}{M} \sum_{m = 1}^M {{\rm {\bf V}}_{mk}^H{{\rm{\bf n}}_m}} $.
\begin{coro}
\vspace*{-0.1cm}
An achievable SE for UE k in the CF XL-MIMO is \cite{9737367}
\begin{equation}
\begin{aligned}
\label{eq:seventeen}{{\rm SE}_k^{(1)}} = {\log _2} \left| {{\rm {\bf I}}_{N_S} + {\rm {\bf E}}_{k,(1)}^H{\bm \Psi} _{k,(1)}^{ - 1}{{\rm {\bf E}}_{k,(1)}}} \right|,
 \end{aligned}
\end{equation}
where $ {{\rm {\bf E}}_{k,(1)}} \triangleq \sqrt p \sum_{m = 1}^M {{\mathbb E}\left\{ {{\rm {\bf V}}_{mk}^H{{\rm {\bf H}}_{mk}}} \right\}}    $ and $  {\bm \Psi} _{k,(1)} \triangleq p\sum_{l = 1}^K {\sum_{m = 1}^M {\sum_{m' = 1}^M {{\mathbb E}\left\{ {{\rm {\bf V}}_{mk}^H{{\rm {\bf H}}_{ml}}{\rm {\bf V}}_{m'l}^H{{\rm {\bf H}}_{m'k}}} \right\}} } }  - {{\rm {\bf E}}_{k,(1)}}{{\rm {\bf E}}_{k,(1)}}^H + \sum_{m = 1}^M {{\mathbb E}\left\{ {{\rm {\bf V}}_{mk}^H{{\rm {\bf n}}_m}{\rm {\bf n}}_m^H{{\rm {\bf V}}_{mk}}} \right\}}  $.
\end{coro}
Note that \eqref{eq:seventeen} can adopt any combining matrix. We can derive the closed form SE expression with MR combining $ {\rm {\bf V}}_{mk}={\rm {\bf H}}_{mk} $ as the following theorem.
\begin{thm}\label{Th_x}
If MR combining is considered, the closed-form SE expression in  the CF XL-MIMO can be derived as
\begin{equation}
\begin{aligned}
{{\rm SE}_k^{(1)}} = {\log _2} \left| {{\rm {\bf I}}_{N_S} + {\rm {\bf E}}_{k,(1)}^H{\bm \Psi} _{k,(1)}^{ - 1}{{\rm {\bf E}}_{k,(1)}}} \right|,
 \end{aligned}
\end{equation}
where $ {{\rm {\bf E}}_{k,(1)}} = \sqrt p \sum_{m = 1}^M {{\rm {\bf Z}}_{mk}} $ and $  {\bm \Psi} _{k,(1)} = p\!\sum_{l = 1}^K \!{\sum_{m = 1}^M \!{\sum_{m' = 1}^M \!{\rm {\bf T}}_{mkm'l}  } }  \!-\! {{\rm {\bf E}}_{k,(1)}}{{\rm {\bf E}}_{k,(1)}}^H \!\!+\!\! \sigma_w^2 \!\sum_{m = 1}^M  \!{{\rm {\bf Z}}_{mk}}  $.
\end{thm}
\begin{IEEEproof}
The proof is given in Appendix A.
\end{IEEEproof}

%%%%%%%%%%%%%%%%%%%%%%%%%%%%%%%%%%%%%%%%%%%%%%%%%%%%%%%%%%%%%%%%%%%%%%%
%-------------------------Small-Cell Network------------------------
%%%%%%%%%%%%%%%%%%%%%%%%%%%%%%%%%%%%%%%%%%%%%%%%%%%%%%%%%%%%%%%%%%%%%%
\subsection{Small-Cell XL-MIMO}
In the small-cell XL-MIMO, the BSs decode the received signals without exchange anything with the CPU. In this case, the signal from one UE is decoded by only one BS.
\begin{coro}
\vspace*{-0.1cm}
An achievable SE for UE k in the small-cell XL-MIMO is
\begin{equation}\label{eq:nineteen}
\begin{aligned}
\!\!\!\!\!\!\!\!  {\rm SE}_k^{(2)}  \!\!= \!\!\!\!\!\!\! \mathop {\max }\limits_{m \in \left\{ {1, \cdots ,M} \right\}} \!  \underbrace{  {\mathbb E}\left\{ {\log _2} \left|  {\rm {\bf I}}_{N_S} \!\! + \!\!   {{\rm {\bf E}}_{mk,(2)}^H{\bm \Psi} _{mk,(2)}^{ - 1}{{\rm {\bf E}}_{mk,(2)}} } \right|\right\}}_{{\rm SE}_{mk}^{(2)}},\!\!
 \end{aligned}
\end{equation}
where $  {\rm {\bf E}}_{mk,(2)}  \triangleq \sqrt p {   {{\rm {\bf V}}_{mk}^H{{\rm {\bf H}}_{mk}}} }   $ and $ {\bm \Psi} _{mk,(2)} \triangleq p\sum_{l = 1,l \neq k}^K { {  {{\rm {\bf V}}_{mk}^H{{\rm {\bf H}}_{ml}}{\rm {\bf H}}_{ml}^H{{\rm {\bf V}}_{mk}}} } }     + \sigma_w^2 {  {{\rm {\bf V}}_{mk}^H{{\rm {\bf V}}_{mk}}} } $. If MR combining is used, the expression of $ SE_{mk} $ in \eqref{eq:nineteen} can be derived as \eqref{eq:twenty}, shown at the top of this page.
\end{coro}

\begin{IEEEproof}
The proof is given in Appendix B.
\end{IEEEproof}

%%%%%%%%%%%%%%%%%%%%%%%%%%%%%%%%%%%%%%%%%%%%%%%%%%%%%%%%%%%%%%%%%%%%%%
%-----------------------numerical results--------------------
%%%%%%%%%%%%%%%%%%%%%%%%%%%%%%%%%%%%%%%%%%%%%%%%%%%%%%%%%%%%%%%%%%%%%
\section{Numerical Results}

In this section, we compare the uplink performance of  the CF XL-MIMO and the small-cell XL-MIMO with MR combining and different antenna spacing.
%Moreover, we investigate the effect of antenna spacing on the performance.

Figure 2 shows the sum SE for two processing schemes over MR combining as a function of $N_{H_r} = N_{V_r}$ with $M=20$, $K=8$, $N_s = N_{H_s} \times N_{V_s} = 36 $ and $\Delta_s = \Delta_r = {\lambda}/{3}$.
The first observation is that the CF XL-MIMO outperforms the small-cell XL-MIMO. And the performance gap between the CF XL-MIMO and the small-cell XL-MIMO increases with the increase of $N_{H_s} = N_{V_s}$.
Besides, the curves generated by Monte Carlo simulations are overlapped by markers ``$\circ$" generated by analytical results in the CF XL-MIMO. This validates the closed-form SE expression derived in Section III. Moreover, we observe that the sum SE reach the maximum values at $N_{H_r} = N_{V_r}=9$, then decrease with the increase of $N_{H_r} = N_{V_r}$.
The reason is that increasing $N_s$ will increase both the interference and desired signal while the performance loss caused by the increase of the interference outweighs the gain in increasing desired signal.

\begin{figure}
\centering
\includegraphics[width=2.7in]{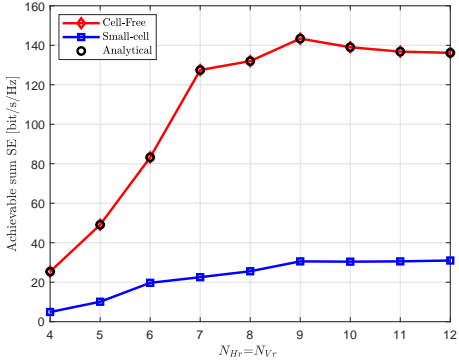}
\caption{Sum SE for two signal processing schemes as a function of $N_{H_r} = N_{V_r}$ with $M=20$, $K=8$, $N_s = N_{H_s} \times N_{V_s} = 36 $ and $\Delta_s = \Delta_r = {\lambda /3}$.}
\end{figure}

\begin{figure}
\centering
\includegraphics[width=2.7in]{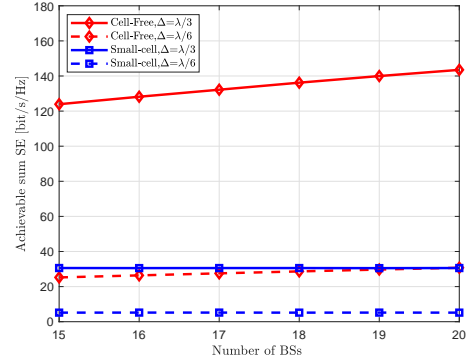}
\caption{Sum SE against the number of BSs for two signal processing schemes over different antenna spacing with $K=8$, $N_s = N_{H_s} \times N_{V_s} = 36 $, and $N_r = N_{H_r} \times N_{V_r} = 81 $.}
\end{figure}

Figure 3 exploits the sum SE against the number of BSs for  two processing schemes over different antenna spacing with $K=8$, $N_s = N_{H_s} \times N_{V_s} = 36 $, and $N_r = N_{H_r} \times N_{V_r} = 81 $.
For the CF XL-MIMO or the small-cell XL-MIMO, we observe that the scheme with $\Delta_s = \Delta_r = {\lambda}/{3}$ performs better than the scheme with $\Delta_s = \Delta_r = {\lambda}/{6}$.
In  the CF XL-MIMO or the small-cell XL-MIMO with $M=20$, compared with the scheme with $\Delta_s = \Delta_r = {\lambda}/{3}$, the scheme with a lower antenna spacing $\Delta_s = \Delta_r = {\lambda}/{6}$ can result in about $ 78.57 \% $ and $ 83.11 \% $ sum SE loss, respectively.
This can be explained by the fact that with the fixed number of antennas, smaller spacing has less surface area and leads to a higher correlation level among patch antennas.
Thus, the SE performance  of the scheme with smaller spacing is worse.
As the number of BSs increases, the sum SE of  the CF XL-MIMO undoubtedly increases while the sum SE of the small-cell XL-MIMO is constant, since the channel model in this paper only considers the small-scale fading and one UE is served by only one BS.

Figure 4 investigates the average SE as a function of the number of UEs $K$ for two processing schemes over different antenna spacing with $M=20$, $N_s = N_{H_s} \times N_{V_s} = 36 $, and $N_r = N_{H_r} \times N_{V_r} = 81 $.
We observe that the performance gap between $\Delta_s = \Delta_r = {\lambda}/{3}$ and $\Delta_s = \Delta_r = {\lambda}/{6}$ becomes smaller with the increase of $K$.
Moreover, we notice that the CF XL-MIMO provides higher SEs than the small-cell XL-MIMO which can be also observed in Figure 2.

\begin{figure}
\centering
\includegraphics[width=2.7in]{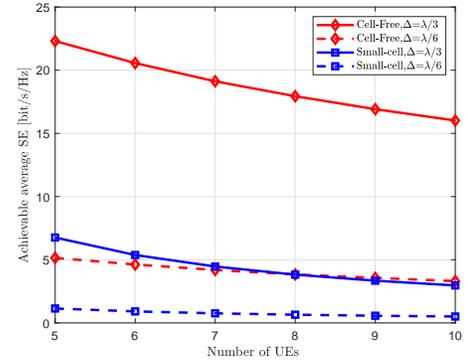}
\caption{Average SE against the number of UEs for two signal processing schemes over different antenna spacing with $M=20$, $N_s = N_{H_s} \times N_{V_s} = 36 $, and $N_r = N_{H_r} \times N_{V_r} = 81 $.}
\end{figure}

\begin{figure}
\centering
\includegraphics[width=2.7in]{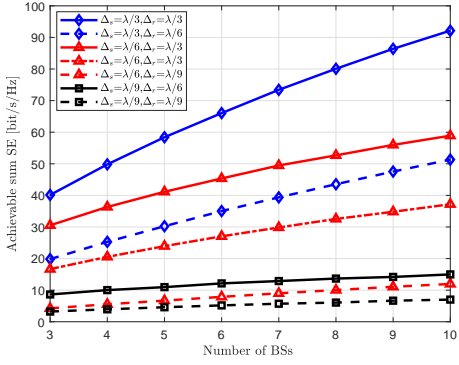}
\caption{Sum SE as a function of the number of BSs for  the distributed XL-MIMO over different antenna spacing with $K=3$, $N_s = N_{H_s} \times N_{V_s} = 144 $, and $N_r = N_{H_r} \times N_{V_r} = 144 $.}
\end{figure}

Figure 5 shows the sum SE as a function of the number of BSs for the
CF XL-MIMO with $K=3$, $N_s = N_{Hs} \times N_{Vs} = 144 $, and $N_r = N_{Hr} \times N_{Vr} = 144 $.
We compare seven cases with different BS antenna spacing $\Delta_r$ and UE antenna spacing $\Delta_s$.
As observed, smaller spacing results in  worse performance.
In fact, the smaller antenna spacing, the higher the correlation level.
Specifically, compared with the case with $\Delta_s = {\lambda}/{6}, \Delta_r = {\lambda}/{6}$,   the case with $\Delta_s = {\lambda}/{3}, \Delta_r = {\lambda}/{6}$ can achieve about $ 138 \%  $ SE improvement, while the case with $\Delta_s = {\lambda}/{6}, \Delta_r = {\lambda}/{3}$ can achieve about $ 158\%  $ SE improvement.
This phenomenon indicates that the antenna spacing of BSs has a greater impact on performance than the antenna spacing of UEs.
The reason is that uplink MR combining can spatially suppress transmit correlation.

%%%%%%%%%%%%%%%%%%%%%%%%%%%%%%%%%%%%%%%%%%%%%%%%%%%%%%%%%%%%%%%%%%%%%%
%----------------------CONCLUSIONS-------------------
%%%%%%%%%%%%%%%%%%%%%%%%%%%%%%%%%%%%%%%%%%%%%%%%%%%%%%%%%%%%%%%%%%%%%
\section{Conclusions }

In this paper, we extend the channel model of single-BS multi-UE XL-MIMO
to the channel model of practical multi-BS multi-UE XL-MIMO.
We investigate the uplink performance of the XL-MIMO system and consider two different signal processing schemes.
Then we derive exact achievable SE expressions for any combining scheme and compute the closed-form SE expression for the CF XL-MIMO with MR combining.
In numerical results, we investigate the impact of antennas spacing and prove that the CF XL-MIMO outperforms the small-cell XL-MIMO.
Moreover, we reveal that  increasing the number of  UE antennas may decrease the SE with MR combining.
We also prove that the decrease of antenna spacing induces stronger correlations.
Considering the impact of polarized antenna arrays and developing novel signal processing schemes based on the CF XL-MIMO channel characteristics could be important topics for future work.

%%%%%%%%%%%%%%%%%%%%%%%%%%%%%%%%%%%%%%%%%%%%%%%%%%%%%%%%%%%%%%%%%%%%%%
%-----------------------equation--------------------
%%%%%%%%%%%%%%%%%%%%%%%%%%%%%%%%%%%%%%%%%%%%%%%%%%%%%%%%%%%%%%%%%%%%%
\begin{figure*}[!t]
\normalsize
\setcounter{equation}{19}
\begin{equation}\label{eq:twenty four}
{{\mathbb E}  \left\{ {{\rm {\bf h}}_{mk,n}^H{{\rm {\bf h}}_{mk,i}}{\rm {\bf h}}_{mk,i}^H{{\rm {\bf h}}_{mk,n'}}} \right\}} \\
 = {\mathbb E}\left\{ {{{\left( {\sum\limits_{{j_1} = 1}^{Ns} {\tilde {\rm {\bf R}}_{mk}^{nj_1}{{\rm{\bf x}}_{j_1}}} } \right)}^H}\left( {\sum\limits_{{j_2} = 1}^{Ns} {\tilde {\rm {\bf R}}_{mk}^{ij_2}{{\rm{\bf x}}_{j_2}}} } \right){{\left( {\sum\limits_{{j_3} = 1}^{Ns} {\tilde {\rm {\bf R}}_{mk}^{ij_3}{{\rm{\bf x}}_{j_3}}} } \right)}^H}\left( {\sum\limits_{{j_4} = 1}^{Ns} {\tilde {\rm {\bf R}}_{mk}^{n'j_4}{{\rm{\bf x}}_{j_4}}} } \right)} \right\}
\end{equation}
\hrulefill
\vspace*{0.1pt}
\end{figure*}
\begin{figure*}[!t]
\normalsize
\setcounter{equation}{20}
\begin{equation}\label{eq:twenty five}
{\mathbb E}\left\{ {\left( {\sum\limits_{j_1 = 1}^{Ns} {{\rm{\bf x}}_{j_1}^H\tilde {\rm {\bf R}}_{mk}^{j_1n}\tilde {\rm {\bf R}}_{mk}^{ij_1}{{\rm{\bf x}}_{j_1}}} } \right){{\left( {\sum\limits_{j_3 = 1}^{Ns} {{\rm{\bf x}}_{j_3}^H\tilde {\rm {\bf R}}_{mk}^{n'j_3}\tilde {\rm {\bf R}}_{mk}^{j_3i}{{\rm{\bf x}}_{j_3}}} } \right)}^H}} \right\}
 =\sum\limits_{j_1 = 1}^{Ns} {\sum\limits_{j_3 = 1}^{Ns} {{\rm{tr}}\left( {\tilde {\rm {\bf R}}_{mk}^{j_1n}\tilde {\rm {\bf R}}_{mk}^{ij_1}} \right){\rm tr}\left( {\tilde {\rm {\bf R}}_{mk}^{n'j_3}\tilde {\rm {\bf R}}_{mk}^{j_3i}} \right)} }
\end{equation}
\hrulefill
\vspace*{0.1pt}
\end{figure*}
\begin{figure*}[!t]
\normalsize
\setcounter{equation}{21}
\begin{equation}\label{eq:twenty six}
{\mathbb E}\left\{ {\left( {\sum\limits_{j_1 = 1}^{Ns} {{\rm{\bf {\bf x}}}_{j_1}^H\tilde {\rm {\bf R}}_{mk}^{j_1n}} } \right)\left( {\sum\limits_{j_2 = 1}^{Ns} {\tilde {\rm {\bf R}}_{mk}^{ij_2}{{\rm{{\bf x}}}_{j_2}}{\rm{\bf x}}_{j_2}^H\tilde {\rm {\bf R}}_{mk}^{j_2i}} } \right)\left( {\sum\limits_{j_1 = 1}^{Ns} {\tilde {\rm {\bf R}}_{mk}^{n'j_1}{{\rm{{\bf x}}}_{j_1}}} } \right)} \right\} = \sum\limits_{j_1 = 1}^{Ns} {\sum\limits_{j_2 = 1}^{Ns} {{\rm{tr}}\left( {\tilde {\rm {\bf R}}_{mk}^{j_1n}\tilde {\rm {\bf R}}_{mk}^{ij_2}\tilde {\rm {\bf R}}_{mk}^{j_2i}\tilde {\rm {\bf R}}_{mk}^{n'j_1}} \right)} }
\end{equation}
\hrulefill
\vspace*{0.1pt}
\end{figure*}
\begin{figure*}[!t]
\normalsize
\setcounter{equation}{22}
\begin{equation}\label{eq:twenty seven}
{\left[ {{{\bm \Gamma} _{mk}^{(2)}}} \right]_{nn'}} = \sum\limits_{j_1 = 1}^{Ns} \sum\limits_{j_2 = 1}^{Ns} \left[  {{\rm{tr}}\left( {\tilde {\rm {\bf R}}_{mk}^{j_1n}\tilde {\rm {\bf R}}_{mk}^{ij_1}} \right){\rm tr}\left( {\tilde {\rm {\bf R}}_{mk}^{n'j_2}\tilde {\rm {\bf R}}_{mk}^{j_2i}} \right)} +  {{\rm{tr}}\left( {\tilde {\rm {\bf R}}_{mk}^{j_1n}\tilde {\rm {\bf R}}_{mk}^{ij_2}\tilde {\rm {\bf R}}_{mk}^{j_2i}\tilde {\rm {\bf R}}_{mk}^{n'j_1}} \right)}  \right]
\end{equation}
\hrulefill
\vspace*{0.1pt}
\end{figure*}
%%%%%%%%%%%%%%%%%%%%%%%%%%%%%%%%%%%%%%%%%%%%%%%%%%%%%%%%%%%%%%%%%%%%%%
%-----------------------equation--------------------
%%%%%%%%%%%%%%%%%%%%%%%%%%%%%%%%%%%%%%%%%%%%%%%%%%%%%%%%%%%%%%%%%%%%%

%%%%%%%%%%%%%%%%%%%%%%%%%%%%%%%%%%%%%%%%%%%%%%%%%%%%%%%%%%%%%%%%%%%%%%
%-----------------------APPENDIX--------------------
%%%%%%%%%%%%%%%%%%%%%%%%%%%%%%%%%%%%%%%%%%%%%%%%%%%%%%%%%%%%%%%%%%%%%
\section*{Appendix}
\subsection{Proof of Theorem 1}
This subsection derive the closed-form SE expression for  the distributed XL-MIMO based on MR combining $ {\rm {\bf V}}_{mk}={\rm {\bf H}}_{mk} $.

We begin with the first term $ {{\rm {\bf E}}_{k,(1)}} = \sqrt p \sum_{m = 1}^M {{\mathbb E}\left\{ {{\rm {\bf V}}_{mk}^H{{\rm {\bf H}}_{mk}}} \right\}}  =  \sqrt p \sum_{m = 1}^M {{\rm {\bf Z}}_{mk}} $, where $ {{\rm {\bf Z}}_{mk}} =  {{\mathbb E}\left\{ {{\rm {\bf V}}_{mk}^H{{\rm {\bf H}}_{mk}}} \right\}} ={{\mathbb E}\left\{ {{\rm {\bf H}}_{mk}^H{{\rm {\bf H}}_{mk}}} \right\}}     \in {\mathbb C}^{N_s\times N_s} $ and the $ ( n,n')$-th element of $ {{\rm {\bf Z}}_{mk}} $ is denoted as $ \left[{{\rm {\bf Z}}_{mk}}\right]_{n,n'}={\mathbb E}\left\{ {{\rm{\bf h}}_{mk,n}^H{{\rm{\bf h}}_{mk,n'}}} \right\} = {\rm tr}\left( {{\rm{\bf R}}_{mk}^{n'n}} \right) $.

Then the second term $ {{\mathbb E}\left\{ {{\rm {\bf V}}_{mk}^H{{\rm {\bf n}}_m}{\rm {\bf n}}_m^H{{\rm {\bf V}}_{mk}}} \right\}}  $ can be computed as
$   {{\mathbb E}\left\{ {{\rm {\bf V}}_{mk}^H{{\rm {\bf n}}_m}{\rm {\bf n}}_m^H{{\rm {\bf V}}_{mk}}} \right\}} =  {{\mathbb E}\left\{ {{\rm {\bf V}}_{mk}^H {{\rm {\bf V}}_{mk}} {{\rm {\bf n}}_m}{\rm {\bf n}}_m^H} \right\}}   = {{\mathbb E}\left\{ {{\rm {\bf H}}_{mk}^H {{\rm {\bf H}}_{mk}} {{\rm {\bf n}}_m}{\rm {\bf n}}_m^H} \right\}} = \sigma_w^2   {{\rm {\bf Z}}_{mk}}  $.

Last but not least, we compute the last term $  {\rm {\bf T}}_{mkm'l} =  {{\mathbb E}\left\{ {{\rm {\bf V}}_{mk}^H{{\rm {\bf H}}_{ml}}{\rm {\bf V}}_{m'l}^H{{\rm {\bf H}}_{m'k}}} \right\}} = {{\mathbb E}\left\{ {{\rm {\bf H}}_{mk}^H{{\rm {\bf H}}_{ml}}{\rm {\bf H}}_{m'l}^H{{\rm {\bf H}}_{m'k}}} \right\}}$ for all possible AP and UE combinations.

{\bf Case 1:} $ m \neq m',l \neq k$

Since $ {\rm {\bf H}}_{mk} $ and $ {\rm {\bf H}}_{ml} $ are independent and both have zero mean, we have $ {{\mathbb E}\left\{ {{\rm {\bf H}}_{mk}^H{{\rm {\bf H}}_{ml}}{\rm {\bf H}}_{m'l}^H{{\rm {\bf H}}_{m'k}}} \right\}} = 0 $.

{\bf Case 2:} $ m \neq m',l = k$

Since $ {\rm {\bf H}}_{mk} $ and $ {\rm {\bf H}}_{m'k} $ are independent, we have $ {{\mathbb E}\left\{ {{\rm {\bf H}}_{mk}^H{{\rm {\bf H}}_{ml}}{\rm {\bf H}}_{m'l}^H{{\rm {\bf H}}_{m'k}}} \right\}} = {{\mathbb E}\left\{ {{\rm {\bf H}}_{mk}^H{{\rm {\bf H}}_{mk}}{\rm {\bf H}}_{m'k}^H{{\rm {\bf H}}_{m'k}}} \right\}} \!\!=\!\! {{\mathbb E}\left\{ {{\rm {\bf H}}_{mk}^H{{\rm {\bf H}}_{mk}} } \right\}} {{\mathbb E}\left\{ { {\rm {\bf H}}_{m'k}^H{{\rm {\bf H}}_{m'k}}} \right\}} =  {\rm {\bf Z}}_{mk}   {\rm {\bf Z}}_{m'k}$.

{\bf Case 3:} $ m = m',l \neq k$

In this case, we define $ {{\bm \Gamma} _{mkl}^{(1)}} \triangleq    {{\mathbb E}\left\{ {{\rm {\bf H}}_{mk}^H{{\rm {\bf H}}_{ml}}{\rm {\bf H}}_{ml}^H{{\rm {\bf H}}_{mk}}} \right\}} \in {\mathbb C}^{N_s\times N_s}   $ with $  {\left[ {{{\bm \Gamma} _{mkl}^{(1)}}} \right]_{nn'}}  =    \sum_{i = 1}^{{N_s}} {{\mathbb E}\left\{ {{\rm {\bf h}}_{mk,n}^H{{\rm {\bf h}}_{ml,i}}{\rm {\bf h}}_{ml,i}^H{{\rm {\bf h}}_{mk,n'}}} \right\}}  $ being $ (n,n') $-th element of $  {{{\bm \Gamma} _{mkl}^{(1)}}}  $. Since $ {\rm {\bf h}}_{mk} $ is independent of $ {\rm {\bf h}}_{ml} $, we have
$
 \!\! {{\mathbb E} \! \left\{ {{\rm {\bf h}}_{mk,n}^H{{\rm {\bf h}}_{ml,i}}{\rm {\bf h}}_{ml,i}^H{{\rm {\bf h}}_{mk,n'}}} \right\}}
=\!{\rm tr}\! \left(    {\mathbb E} \! \left\{ {{\rm {\bf h}}_{ml,i}{{\rm {\bf h}}_{ml,i}^H} }   \right\} \!  {\mathbb E} \! \left\{ {{\rm {\bf h}}_{mk,n'}{{\rm {\bf h}}_{mk,n}^H} }   \right\}     \right) \!=\! {\rm tr}\! \left(  {{\rm {\bf R}}_{ml}^{ii}{\rm {\bf R}}_{mk}^{n'n}}    \right).
$

So, we can obtain
$
{\left[ {{{\bm \Gamma} _{mkl}^{(1)}}} \right]_{nn'}}  =    \sum_{i = 1}^{{N_s}}
{\rm tr} \left(  {{\rm {\bf R}}_{ml}^{ii}{\rm {\bf R}}_{mk}^{n'n}}    \right)
$.

{\bf Case 4:} $ m = m',l = k$

We first define $  {{\bm \Gamma} _{mk}^{(2)}} \triangleq  {{\mathbb E}\left\{ {{\rm {\bf H}}_{mk}^H{{\rm {\bf H}}_{mk}}{\rm {\bf H}}_{mk}^H{{\rm {\bf H}}_{mk}}} \right\}} \in {\mathbb C}^{N_s\times N_s}  $ whose $ (n,n') $-th element is
\begin{equation}
\begin{aligned}\label{eq:twenty three}
\setcounter{equation}{19}
{\left[ {{{\bm \Gamma} _{mk}^{(2)}}} \right]_{nn'}} = \sum_{i = 1}^{{N_s}} {{\mathbb E}  \left\{ {{\rm {\bf h}}_{mk,n}^H{{\rm {\bf h}}_{mk,i}}{\rm {\bf h}}_{mk,i}^H{{\rm {\bf h}}_{mk,n'}}} \right\}}.
%\\
%&= \sum_{i = 1}^{{N_s}}{\rm tr} ( {{\mathbb E}  \left\{ {{{\rm {\bf h}}_{mk,i}}{\rm {\bf h}}_{mk,i}^H{{\rm {\bf h}}_{mk,n'}}}{\rm {\bf h}}_{mk,n}^H \right\}}  ).
 \end{aligned}
\end{equation}

According to \cite{9737367}, we can then rewrite $  {{\rm {\bf h}}_{mk,i}}  $ and $ {{\rm {\bf h}}_{mk,  \{n,n' \}}}  $ as  $   {{\rm {\bf h}}_{mk,i}}  = \sum_{j = 1}^{Ns} {\tilde {\rm {\bf R}}_{mk}^{ij}{{\rm {\bf x}}_j}}    $, $   {{\rm {\bf h}}_{mk,n}}  = \sum_{j = 1}^{Ns} {\tilde {\rm {\bf R}}_{mk}^{nj}{{\rm {\bf x}}_j}}    $ and $   {{\rm {\bf h}}_{mk,n'}}  = \sum_{j = 1}^{Ns} {\tilde {\rm {\bf R}}_{mk}^{n'j}{{\rm {\bf x}}_j}}    $, where $ \tilde {\rm {\bf R}}_{mk}^{nj} $ is the $ (n,j) $-submatrix of $ {\tilde {\rm {\bf R}}_{mk}^{1/2}} $ and $ {{\rm {\bf x}}_j} \sim {\mathcal CN}\left(  0 , {\rm {\bf I}}_{N_r} \right)$.
So we can obtain \eqref{eq:twenty four}, shown at the top of next page.
According to \cite{9737367}, \eqref{eq:twenty four} will be non-zero only for the case of $  j_1=j_2,j_3=j_4  $ and $  j_1=j_4,j_2=j_3  $. If $  j_1=j_2,j_3=j_4  $, \eqref{eq:twenty four} can be rewritten as \eqref{eq:twenty five}. If $  j_1=j_4,j_2=j_3  $, \eqref{eq:twenty four} can be rewritten as \eqref{eq:twenty six}.

Plugging \eqref{eq:twenty four}, \eqref{eq:twenty five} and \eqref{eq:twenty six} into (19), we can obtain \eqref{eq:twenty seven} to finish the proof.

\subsection{Proof of Corollary 2}

Following similar steps from \cite{9737367}, we have
\begin{equation}
\setcounter{equation}{24}
\label{eq:twenty sevennn}
{\bf {\emph I}} \left( {{{\rm{{\bf x}}}_k};{{{\rm{\check {\bf x}}}}_{mk}},{{\rm{{\bf H}}}_{mk}}} \right) = h\left( {{\rm{{\bf x}}}_{k}|{{\rm{{\bf H}}}_{mk}}} \right) - h\left( {{{\rm{{\bf x}}}_k}|{{{\rm{\check {\bf x}}}}_{mk}},{{\rm{{\bf H}}}_{mk}}} \right),
\end{equation}
where $ h\left( \cdot \right) $ denotes the differential entropy. And we have
\begin{equation}
\label{eq:twenty nine}
h\left(  {{\rm{{\bf x}}}_k}\right) = {\log _2}\left| {\pi e{{\rm {\bf I}}_{{N_S}}}} \right|.
\end{equation}
Then, the estimate of $ {{\rm{{\bf x}}}_k} $ at BS $m$ with $ {{{\rm{\check {\bf x}}}}_{mk}} $ and $ {{\rm{{\bf H}}}_{mk}} $ is
$
\!\!\!{{{\rm{\bar {\bf x}}}}_{mk}} = {\mathbb E}   \left\{ {\sqrt p {\rm {\bf V}}_{mk}^H{{\rm {\bf H}}_{mk}}|{{\rm{{\bf H}}}_{mk}}} \right\}{\mathbb E}     {\left\{ {{{{\rm{\check {\bf x}}}}_{mk}}{\rm{\check {\bf x}}}_{mk}^H|{{\rm{{\bf H}}}_{mk}}} \right\}^{ - 1}}{{{\rm{\check {\bf x}}}}_{mk}}
 = {{\rm {\bf {E}}}_{mk,(2)}\tilde {\bf \Psi} _{k,(2)}^{ - 1}{{{\rm{\check { \bf x}}}}_{mk}}}
$,
where $ {{\rm {\bf {E}}}_{mk,(2)}} = {\mathbb E}   \left\{ {\sqrt p {\rm {\bf V}}_{mk}^H{{\rm {\bf H}}_{mk}}|{{\rm{{\bf H}}}_{mk}}} \right\} = \sqrt p {\rm {\bf V}}_{mk}^H{{\rm {\bf H}}_{mk}} $, $  \tilde {\bf \Psi} _{k,(2)} = {\mathbb E}     {\left\{ {{{{\rm{\check {\bf x}}}}_{mk}}{\rm{\check {\bf x}}}_{mk}^H|{{\rm{{\bf H}}}_{mk}}} \right\}} = {\rm {\bf V}}_{mk}^H\left( {p\sum_{l = 1}^K {{{\rm {\bf H}}_{ml}}{\rm {\bf H}}_{ml}^H}  + \sigma_w^2{\rm {\bf I}}_{N_r}} \right){V_{mk}}$.
The estimation error of $ {{\rm{{\bf x}}}_k} $ is denote by $ \tilde {\rm{{\bf x}} }_{mk} = {{\rm{{\bf x}}}_{mk}} -{{{\rm{\check{\bf x}}}}_{mk}} $, then $  h\left( {{{\rm{{\bf x}}}_k}|{{{\rm{\check {\bf x}}}}_{mk}},{{\rm{{\bf H}}}_{mk}}} \right) $ is upper bounded by
\begin{equation}
\begin{aligned}
\label{eq:thirty-one}
h\left( {{{\rm{{\bf x}}}_k}|{{{\rm{\check {\bf x}}}}_{mk}},{{\rm{{\bf H}}}_{mk}}} \right)  \le  {\mathbb E}    \left\{ {{{\log }_2}\left| {\pi e{\mathbb E}\left\{ {{\tilde {\rm{{\bf x}} }_{mk}}\tilde {\rm{{\bf x}} }_{mk}^H|{{\rm {\bf H}}_{mk}}} \right\}} \right|} \right\}\\
 = {\mathbb E} \left\{ {{{\log }_2}\left| {\pi e\left( {{\rm {\bf I}}_{N_S} - {{\rm {\bf {E}}}_{mk,(2)}}\tilde {\bf \Psi} _{mk,(2)}^{ - 1}{\rm {\bf {E}}}_{mk,(2)}^H} \right)} \right|} \right\}.
\end{aligned}
\end{equation}
Plugging \eqref{eq:twenty nine} and \eqref{eq:thirty-one} into (24), we have
$
{\bf {\emph I}} \left( {{{\rm{{\bf x}}}_k};{{{\rm{\check {\bf x}}}}_{mk}},{{\rm{{\bf H}}}_{mk}}} \right) \!\ge\! {\mathbb E} \left\{ {{{\log }_2}\left| { {{\rm {\bf I}}_{N_S} \!\!+\! {{\rm {\bf {E}}}_{mk,(2)}^H} {\bf \Psi} _{mk,(2)}^{ - 1}{\rm {\bf {E}}}_{mk,(2)}}} \right|} \right\}
$,
where $ {\bf \Psi} _{mk,(2)} = {\rm {\bf V}}_{mk}^H\left( {p\sum_{l = 1,l \leq k}^K {{{\rm {\bf H}}_{ml}}{\rm {\bf H}}_{ml}^H}  + \sigma_w^2{\rm {\bf I}}_{N_r}} \right){V_{mk}} $. To finish the proof,  an achievable SE for UE $k$ can be derived as \eqref{eq:nineteen}.

\bibliographystyle{IEEEtran}
\bibliography{IEEEabrv,Ref}

% Generated by IEEEtran.bst, version: 1.14 (2015/08/26)
\begin{thebibliography}{10}
\providecommand{\url}[1]{#1}
\csname url@samestyle\endcsname
\providecommand{\newblock}{\relax}
\providecommand{\bibinfo}[2]{#2}
\providecommand{\BIBentrySTDinterwordspacing}{\spaceskip=0pt\relax}
\providecommand{\BIBentryALTinterwordstretchfactor}{4}
\providecommand{\BIBentryALTinterwordspacing}{\spaceskip=\fontdimen2\font plus
\BIBentryALTinterwordstretchfactor\fontdimen3\font minus
  \fontdimen4\font\relax}
\providecommand{\BIBforeignlanguage}[2]{{%
\expandafter\ifx\csname l@#1\endcsname\relax
\typeout{** WARNING: IEEEtran.bst: No hyphenation pattern has been}%
\typeout{** loaded for the language `#1'. Using the pattern for}%
\typeout{** the default language instead.}%
\else
\language=\csname l@#1\endcsname
\fi
#2}}
\providecommand{\BIBdecl}{\relax}
\BIBdecl

\bibitem{bjornson2019massive}
E.~Bj{\"o}rnson, L.~Sanguinetti, H.~Wymeersch, J.~Hoydis, and T.~L. Marzetta,
  ``Massive {MIMO} is a reality{---}what is next? {Five} promising research
  directions for antenna arrays,'' \emph{Digit. Signal Process.}, vol.~94, pp.
  3--20, Nov. 2019.

\bibitem{9113273}
J.~Zhang, E.~Bj{\"o}rnson, M.~Matthaiou, D.~W.~K. Ng, H.~Yang, and D.~J. Love,
  ``Prospective multiple antenna technologies for beyond {5G},'' \emph{IEEE J.
  Sel. Areas Commun.}, vol.~38, no.~8, pp. 1637--1660, Aug. 2020.

\bibitem{https://doi.org/10.48550/arxiv.2209.12131}
Z.~Wang, J.~Zhang, H.~Du, W.~E.~I. Sha, B.~Ai, D.~Niyato, and M.~Debbah,
  ``Extremely large-scale {MIMO}: Fundamentals, {Challenges}, {Solutions}, and
  {Future} {D}irections,'' {\it arXiv:2209.12131}, 2022.

\bibitem{9903389}
M.~Cui, Z.~Wu, Y.~Lu, X.~Wei, and L.~Dai, ``Near-field communications for {6G}:
  Fundamentals, challenges, potentials, and future directions,'' \emph{IEEE
  Commun. Mag.}, pp. 1--7, to appear, 2022.

\bibitem{https://doi.org/10.48550/arxiv.2209.03082}
P.~Ramezani and E.~Bj{\"o}rnson, ``Near-field beamforming and multiplexing
  using extremely large aperture arrays,'' {\it arxiv:2209.03082}, 2022.

\bibitem{9650519}
S.~S.~A. Yuan, Z.~He, X.~Chen, C.~Huang, and W.~E.~I. Sha, ``Electromagnetic
  effective degree of freedom of an {MIMO} system in free space,'' \emph{IEEE
  Antennas Wireless Propag. Lett.}, vol.~21, no.~3, pp. 446--450, Mar. 2022.

\bibitem{9136592}
C.~Huang, S.~Hu, G.~C. Alexandropoulos, A.~Zappone, C.~Yuen, R.~Zhang, M.~D.
  Renzo, and M.~Debbah, ``Holographic {MIMO} surfaces for {6G} wireless
  networks: {Opportunities}, challenges, and trends,'' \emph{IEEE Wireless
  Commun.}, vol.~27, no.~5, pp. 118--125, Oct. 2020.

\bibitem{9443506}
A.~Pizzo, T.~Marzetta, and L.~Sanguinetti, ``Holographic {MIMO} communications
  under spatially-stationary scattering,'' in \emph{2020 54th Asilomar Conf.
  Signals Syst. Comput.}, 2020, pp. 702--706.

\bibitem{9110848}
A.~Pizzo, T.~L. Marzetta, and L.~Sanguinetti, ``Spatially-stationary model for
  holographic {MIMO} small-scale fading,'' \emph{IEEE J. Sel. Areas Commun.},
  vol.~38, no.~9, pp. 1964--1979, Sept. 2020.

\bibitem{9716880}
{\"O}.~T. Demir, E.~Bj{\"o}rnson, and L.~Sanguinetti, ``Channel modeling and
  channel estimation for holographic massive {MIMO} with planar arrays,''
  \emph{IEEE Wireless Commun. Lett.}, vol.~11, no.~5, pp. 997--1001, 2022.

\bibitem{9724113}
A.~Pizzo, L.~Sanguinetti, and T.~L. Marzetta, ``Fourier plane-wave series
  expansion for holographic {MIMO} communications,'' \emph{IEEE Trans. Wirel.
  Commun.}, pp. 1--1, Mar. 2022.

\bibitem{9779586}
L.~Wei, C.~Huang, G.~Alexandropoulus, W.~E. Sha, Z.~Zhang, M.~Debbah, and
  C.~Yuen, ``Multi-user holographic {MIMO} surfaces: {Channel} modeling and
  spectral efficiency analysis,'' \emph{IEEE J. Sel. Topics Signal Process.},
  pp. 1--1, to appear, 2022.

\bibitem{[1]}
J.~Zhang, S.~Chen, Y.~Lin, J.~Zheng, B.~Ai, and L.~Hanzo, ``Cell-free massive
  {MIMO}: {A} new next-generation paradigm,'' \emph{IEEE Access}, vol.~7, pp.
  99\,878--99\,888, Jul. 2019.

\bibitem{9174860}
S.~Chen, J.~Zhang, E.~Bj{\"o}rnson, J.~Zhang, and B.~Ai, ``Structured massive
  access for scalable cell-free massive {MIMO} systems,'' \emph{IEEE J. Sel.
  Areas Commun.}, vol.~39, no.~4, pp. 1086--1100, Apr. 2021.

\bibitem{9737367}
Z.~Wang, J.~Zhang, B.~Ai, C.~Yuen, and M.~Debbah, ``Uplink performance of
  cell-free massive {MIMO} with multi-antenna users over jointly-correlated
  {Rayleigh} fading channels,'' \emph{IEEE Trans. Wireless Commun.}, vol.~21,
  no.~9, pp. 7391--7406, Sep. 2022.

\end{thebibliography}

\end{document}